\documentclass[10pt,letterpaper, comsoc]{IEEEtran}

\usepackage{graphicx}
\usepackage{amssymb}
\usepackage{amsmath}
\usepackage{amsfonts}
\usepackage{dsfont}
\usepackage{cite}
\usepackage{url}
\usepackage{enumerate}
\usepackage{color}
\usepackage{epstopdf}
\usepackage{array}
\usepackage{tabularx}
\usepackage{balance}
\usepackage{algpseudocode}
\usepackage{algorithm}
\usepackage{multirow}

\allowdisplaybreaks[1]

\makeatletter
\def\footnoterule{\relax%
  \kern-5pt
  \hbox to \columnwidth{\hfill\vrule width \columnwidth height 0.4pt\hfill}
  \kern4.6pt}
\makeatother

\begin{document}


\title{Low-Complexity Multiuser QAM  \\ Detection   for  Uplink   1-bit  Massive MIMO}
\author{
         Panos N. Alevizos,   \IEEEmembership{Student Member, IEEE}
%
}


\maketitle
\begin{abstract}

This work    studies multiuser detection for one-bit massive   multiple-input multiple-output (MIMO) systems
in order to diminish the power consumption at the   base station  (BS). 
A   low-complexity near-maximum-likelihood (nML) multiuser  detection
algorithm  is designed, assuming that
 each BS  antenna port is connected with
a pair of single-bit resolution  analog-to-digital converters (ADCs)  
and  each  user equipment (UE)
 transmits symbols from a quadrature amplitude modulation (QAM) 
constellation.
   First, a novel convex  program   is formulated  as a convex surrogate of the ML detector and subsequently solved 
through an accelerated first-order  method.  Then, the  solution of the convex optimization problem
is  harnessed to solve a  refined combinatorial problem with reduced 
search space, requiring non-exponential complexity on the number of the UEs.
Judicious simulation study corroborates the efficacy of the  resulting two-phase detection algorithm. 
 The proposed two-phase algorithm can achieve  symbol error rate (SER) performance close to the ML detector,  with
significantly reduced computation cost compared to the nML detection schemes in prior art.
\end{abstract}


\IEEEpeerreviewmaketitle

\section{Introduction}

Massive multiple-input multiple-output (MIMO) systems  
in conjunction with single-bit resolution  analog-to-digital converters (ADCs)
will be a promising cost-efficient solution for
future green cellular networks that support wide bandwidths.
In addition to the above,
as the in-phase and the quadrature components of the
 continuous-valued received samples are
quantized separately using one-bit ADCs (i.e., zero-threshold comparators)
the resulting   hardware complexity at the base station (BS) can be sustained to ultra-low levels.

Relevant papers in \cite{WaLiWa:14, JaCDuCoGuSt:15, ChMoHe:16, RisPerLar:14} offer the current perspective
of uplink massive MIMO with one-bit resolution ADCs.
Work in \cite{WaLiWa:14} designed a low-complexity  message-passing
one-bit multiuser detector for quadrature phase-shift keying (QPSK) alphabets at the user equipment devices (UEs),
demonstrating performance close to the linear minimum mean-squared error (MMSE) 
  detector with reduced computational cost.
Subsequent work   \cite{JaCDuCoGuSt:15} offered
  throughput analysis of one-bit multiuser linear detectors in  
uplink  massive MIMO \cite{RisPerLar:14}, quantifying also the impact of imperfect 
channel state information (CSI) at the BS.
Near-maximum-likelihood (nML) detection  
is proposed in~\cite{ChMoHe:16} using a two-stage procedure.

In this paper, we focus on uplink multiuser massive MIMO
systems  with single-bit ADCs at the BS, assuming that the UEs transmit 
symbols from a  square quadrature amplitude modulation (QAM) constellation.
QAM  is the dominating modulation scheme in current 
Long-Term Evolution-Advanced Pro (LTE-A Pro) and future cellular networks \cite{DahParkSk:16}.
The proposed detector is divided in two phases. In the first phase,
a novel convex optimization formulation is proposed, standing as a 
convex surrogate of the ML detection rule; the latter is 
optimal in terms of symbol error rate (SER) but requires  exponential computational cost  on the number of UEs
and a   huge  number of memory resources at the BS.
 The relaxed convex program
is solved through an 
accelerated projected gradient method with adaptive restart, achieving
close to the optimal convergence rate. 
In the second phase of the algorithm, the solution of the   convex program is harnessed
to identify the less-reliable UE symbols  and refine their decision estimates
 via a   combinatorial problem with reduced 
search space.
The resulting   two-phase detector does not require exponential  computational cost on the number of the UEs.
Thorough simulation study  demonstrates that the proposed detector achieves
similar SER performance with the ML detector, and at the same time, significantly reduces the
 computational cost compared to the nML detection schemes in prior art.

\emph{Notation}:
Notation $\mathds{R}$, $\mathds{R}_+$, and $\mathds{C}$, stands for the set of real, non-negative, and complex numbers, respectively.
Nonbold lower-case letters (e.g., $x$) will stand for variables.
Vectors and matrices will be denoted by lower-case (e.g., $\mathbf{x}$) 
and capital (e.g., $\mathbf{A}$), respectively, bold characters.
Symbols $(\cdot)^{\top}$ and  $(\cdot)^{\mathsf{H}}$  denote the transpose  and the conjugate transpose of a vector or matrix, respectively.
$\mathbf{0}_N$ ($\mathbf{1}_N$) and $\mathbf{I}_N$ denote the $N$-dimensional 
all-zeros (all-ones) vector and the $N\times N$ identity matrix. 
$\mathcal{CN}( \boldsymbol{\mu}, \boldsymbol{\Sigma})$ denotes the proper  complex Gaussian
distribution while  $\mathcal{N}( \boldsymbol{\mu}, \boldsymbol{\Sigma})$ denotes the (real) Gaussian
distribution.
 $\bigotimes_{i=1}^N \mathcal{A}_i$ denotes the $N$-fold Cartesian product
of sets $\{\mathcal{A}_i\}_{i=1}^N$.

\section{System Model and Problem Statement}
\label{sec:System model}

We consider an uplink system consisting of a  BS, equipped with $M$ antennas.
  The BS serves $K$ UEs, where $M\gg K$. 
For a single channel use, the received signal at  
the BS, $\mathbf{y} \in \mathds{C}^{M}$,
  is given by 
\begin{equation}
 \mathbf{y} = \mathbf{H} \, \mathbf{P}\, \mathbf{x} + \mathbf{n} =   \sum_{k=1}^K \sqrt{p_k} \,\mathbf{h}_{k} \, x_k  + \mathbf{n},
 \label{signal_model_complex}
\end{equation}
where  $p_k$ is the transmit power of the $k$th UE, $k \in \{1,2,\ldots,K\}$, 
$\mathbf{H} = \left[\mathbf{h}_{1}  \, \mathbf{h}_{2}  \ldots\,  \mathbf{h}_{K} \right] 
\in \mathds{C}^{M\times K}$ is the compound uplink channel matrix consisting of uplink  channel vectors $ \mathbf{h}_{k} \in \mathds{C}^{M}$
from the $k$th UE to the BS, 
$k \in \{1,2,\ldots,K \}$. 
Matrix $\mathbf{P}   \in \mathds{R}_+^{K\times K}$ is a diagonal matrix, whose diagonal elements
comprise of $\{\sqrt{p_k}\}_{k=1}^K$;
$\mathbf{n} \sim \mathcal{CN}(\mathbf{0}_{M}, \sigma^2 \mathbf{I}_M)$
is additive complex Gaussian noise at the  BS of variance $\sigma^2$, while
vector  $\mathbf{x} = [x_{1}\,x_{2}\,\ldots \, x_{K} ]^{\top} \in \mathds{C}^{K}$
 comprises of  the $K$ UEs' transmitted symbols. 
  Each $x_{k}$ belongs to a normalized square $\mathtt{Q}$-QAM   constellation $\mathcal{X}$, i.e.,
  vector $\mathbf{x}$  satisfies $\mathbb{E} [\mathbf{x} ] = \mathbf{0}_K$ and  $\mathbb{E} [\mathbf{x} \, \mathbf{x}^{\mathsf{H}}] = \mathbf{I}_K$.
For that case, $\sqrt{\mathtt{Q}}$ is an integer and $\mathcal{X} 
\triangleq \{ x_{\rm I} + \mathsf{j}x_{\rm Q}: x_{\rm I},x_{\rm Q} \in \mathcal{S} \}$,
where
$
\mathcal{S} \triangleq  \left\{ \sqrt{\frac{3}{2(\mathtt{Q}-1)}}(2q - 1 - \sqrt{\mathtt{Q}})\right\}_{q=1}^{\sqrt{\mathtt{Q}}}
 \label{eq:sqrtM_PAM}
$
is the constellation of  $\sqrt{\mathtt{Q}}$-PAM.
 Each wireless link is subject to Rayleigh small-scale fading, i.e.,  channel vectors  $ \mathbf{h}_{k} \sim \mathcal{CN}(\mathbf{0}_M,
v_{k}^2 \, \mathbf{I}_M)$, where $v_{k}^2$ is the corresponding distance-dependent wireless channel variance.
The resulting   signal-to-noise ratio (SNR) for UE $k$ is
${\rm SNR}_{k} \triangleq \frac{p_k\, v_k^2}{\sigma^2}$.

For a simplified exposition,    the signal model in  Eq.~\eqref{signal_model_complex} is transformed to the real domain  
  as follows
\begin{equation}
 \mathbf{r} = \mathbf{G} \, \mathbf{s} + \mathbf{w},
 \label{signal_model_real}
\end{equation}
 where   $\mathbf{r} \triangleq  \! 
\left[\begin{smallmatrix} 
\Re\{ \mathbf{y}\} \\ \Im\{ \mathbf{y}\} \end{smallmatrix} \! \right] $, 
$\mathbf{G} \! \triangleq \! \left[\begin{smallmatrix} 
\Re\{ \mathbf{H}\, \mathbf{P}\} ~-\Im\{ \mathbf{H}\, \mathbf{P} \} \\  \Im\{ \mathbf{H}\, \mathbf{P}\} ~~\Re\{\mathbf{H}\, \mathbf{P}\} 
\end{smallmatrix} \right] $,  $\mathbf{s} \! \triangleq\! 
\left[\begin{smallmatrix} 
\Re\{ \mathbf{x}\}  \\ \Im\{ \mathbf{x}\} \end{smallmatrix}  \right] $, 
and $ \mathbf{w} \triangleq  \left[\begin{smallmatrix} \Re\{ \mathbf{n}\} \\ \Im\{  \mathbf{n}\} \end{smallmatrix}\! \right] $.
Note that $\mathbf{r} , \mathbf{w} \in \mathds{R}^{2M}$, $\mathbf{G} \in \mathds{R}^{2M \times 2K}$,
while  each element of $\mathbf{s}$ belongs to a $\sqrt{\mathtt{Q}}$-PAM constellation, i.e., $\mathbf{s} \in \mathcal{S}^{2K}$.

BS applies one-bit quantization on the signal $\mathbf{r}$ and forms vector
 $\mathbf{b} = [b_{1}\, b_{2}\, \ldots b_{2M}]^{\top} = \mathsf{sign}(\mathbf{r}) \in \{\pm 1\}^{2M}$, where $\mathsf{sign}(\cdot)$
 is the sign operator applied component-wise. The objective at the BS 
 is to detect $\mathbf{s}$, i.e., the transmitted symbol sequence from the UEs in the cell,
 using only  the one-bit-quantized noisy measurements $\mathbf{b} $.
The noise  vector in~\eqref{signal_model_real} satisfies
$ \mathbf{w} \sim \mathcal{N}\!\left(\mathbf{0}_{2M}, \frac{\sigma^2}{2} \, \mathbf{I}_{2M}\right)$,
and thus, with the compound uplink channel matrix $\mathbf{G} = [\mathbf{g}_{1}\, \mathbf{g}_{2} \, \ldots \, \mathbf{g}_{2M}]^{\top}$
 available, the received vector offers the following statistics
 $\mathbf{r} \sim \mathcal{N}\!\left( \mathbf{G} \, \mathbf{s}, \frac{\sigma^2}{2} \, \mathbf{I}_{2M}\right)$.
Each element of $\mathbf{b}$, $b_m$, follows Bernouli distribution with  $\mathbb{P}(b_m = 1) =  
\mathsf{Q}\!\left(- \frac{\sqrt{ 2}}{ \sigma} \, \mathbf{g}_m^{\top} \mathbf{s}\right)$,
where $\mathsf{Q}(x) = \frac{1}{\sqrt{2 \pi}}\int_{x}^{\infty} \mathsf{e}^{-t^2 /2} \mathsf{d} t$ is the well known Q-function.
 Using similar reasoning with \cite[Eq.~(4)]{TsaJaSidOtt:13}, it follows that the SER-optimal  ML
detector can be expressed as 
\begin{equation}
\widehat{\mathbf{s}}^{\rm ML} = \arg \min_{\mathbf{s} \in  \mathcal{S}^{2K}} \left \{  -\sum_{m=1}^{2M} \mathsf{ln}\,\mathsf{Q}\left( -  \sqrt{\gamma}\,
b_{m} \, \mathbf{g}_{m}^{\top} \,  \mathbf{s} \right) \right \},
\label{eq:ML_detector}
\end{equation}
where $ \gamma = 2/ \sigma^2$.
The complexity to calculate the sequence $\widehat{\mathbf{s}}^{\rm ML}$ in Eq.~\eqref{eq:ML_detector} scales 
as $\mathcal{O}(\sqrt{\mathtt{Q}}^{2K} M\,K) = \mathcal{O}(\mathtt{Q}^{K} M\,K)$, which is exponential on the number of UEs. 

\section{Proposed Near SER-optimal   Detector}
\label{sec:proposed_detector}

In this section  a two-phase detection algorithm is proposed in order to seek an approximate solution to
the ML  detector in~\eqref{eq:ML_detector}, which requires exponential computational cost.

\subsection{Phase I: Relaxation and Projection}
\label{subsec:phase_I_detection}

In the first phase  (phase I), a convex surrogate of optimization problem~\eqref{eq:ML_detector}
is formulated. Specifically, since each element of $\mathbf{s}$, ${s}_{n}$, belongs to a $\sqrt{\mathtt{Q}}$-PAM constellation,
we relax  constraint $\mathbf{s} \in  \mathcal{S}^{2K}$ to
$ |{s}_{n}| \leq \sqrt{\frac{3}{2( {\mathtt{Q}}-1)}}(  \sqrt{\mathtt{Q}} - 1) $,  $n=1,2,\ldots, 2K$.
Hence, the proposed convex relaxation 
version of~\eqref{eq:ML_detector} is expressed as  
 \begin{subequations}\label{eq:phaseI_proposed_detector}
\begin{align}
 \underset{\mathbf{s} \in  \mathds{R}^{2K}}{\rm minimize}  &~ -\sum_{m=1}^{2M} \mathsf{ln}\,\mathsf{Q}\left( -  \sqrt{\gamma}\,
b_{m} \, \mathbf{g}_{m}^{\top} \,  \mathbf{s} \right)  \label{eq:phaseI_proposed_detector_objective}
\\
\text{subject to} & ~|{s}_{n}| \leq \sqrt{\frac{3( \sqrt{\mathtt{Q}} - 1)^2}{2( {\mathtt{Q}}-1)}}  ,  ~ n=1,2,\ldots, 2K. \label{eq:phaseI_proposed_detector_constraints1} 
\end{align}
 \end{subequations}
Note that  function   $\mathsf{f}(\mathbf{s} ) \triangleq  -\sum_{m=1}^{2M} \mathsf{ln}\,\mathsf{Q}\left( - \sqrt{\gamma}\,
b_{m} \, \mathbf{g}_{m}^{\top} \,  \mathbf{s} \right) $ is a convex function of $\mathbf{s} \in \mathds{R}^{2K}$
as a composition of an affine function with a convex  increasing function $-\mathsf{ln}\,\mathsf{Q}(x)$ \cite[p.~84]{BoydVand:04}.
 The set of  constraints in Eq.~\eqref{eq:phaseI_proposed_detector_constraints1}
 is denoted as $\mathcal{B}$, i.e., $\mathcal{B}
\triangleq \left \{\mathbf{s} \in  \mathds{R}^{2K} :  |{s}_{n}| \leq  \sqrt{\frac{3( \sqrt{\mathtt{Q}} - 1)^2}{2( {\mathtt{Q}}-1)}} ,~n=1,2,\ldots, 2K \right\}$,
forming a  box on $\mathds{R}^{2K}$, which is a convex set.
Thus, the problem in~\eqref{eq:phaseI_proposed_detector} is a convex optimization problem \cite{BoydVand:04},
which can be solved either with gradient- or Newton-based iterative algorithms.

In this work,   the optimal solution of   problem~\eqref{eq:phaseI_proposed_detector} is calculated 
through an accelerated  projected gradient method  exploiting the
smoothness  of the objective function  (i.e., continuously differentiable objective) \cite{Bub:15}.
First, the gradient     of $ \mathsf{f}(\cdot )$
 is calculated as \cite{AlFuSidYaBl:18}
 \begin{equation}
\nabla  \mathsf{f}(\mathbf{s})  =  - \sum_{m = 1}^{2 M}
\frac{ \sqrt{\gamma}\,  b_{m}\, \mathsf{e}^{- \frac{\gamma  \left(\mathbf{g}_{m}^{\top}\mathbf{s} \right)^2}{2} } }
{\sqrt{2 \,\pi }   \,   \mathsf{Q}\!\left( - \sqrt{\gamma} \,  b_{m} \, \mathbf{g}_{m}^{\top}\mathbf{s}\right) } \mathbf{g}_{m}
   \label{eq:_gradient_h}  .
\end{equation}
Then,  we need to evaluate an
  upper bound for local smoothness parameter of function $\mathsf{f}(\mathbf{s})$ at any
$\mathbf{s} \in  \mathds{R}^{2K}$, which through the use 
of Cauchy-Swartz inequality for matrix norms, can be obtained
as \cite{AlFuSidYaBl:18}:
$\| \nabla^2  \mathsf{f}(\mathbf{s}) \|_2 \leq  \|  \mathbf{G} \|_2^2  \,  \|\mathbf{d}(\mathbf{s})\|_{\infty}
\triangleq  \mathsf{L}(\mathbf{s}) ,~\forall \mathbf{s} \in \mathds{R}^{2K} , \label{eq:bound_smoothness_h}
$
where the elements of vector function $\mathbf{d}(\mathbf{s})$ are given by
\begin{equation}
  {d}_{m}(\mathbf{s}) \! =   \frac{  \gamma \, \mathsf{e}^{- \gamma \left(\mathbf{g}_{m}^{\top}\mathbf{s} \right)^2}}
{  2\,   \pi   \left[\mathsf{Q}\!\left( - \sqrt{\gamma} \, b_{m} \, \mathbf{g}_{m}^{\top}\mathbf{s} \right)  \right]^2}
 + 
 \frac{  \gamma^{\frac{3}{2}}  b_{m} \,( \mathbf{g}_{m}^{\top}\mathbf{s}) \, \mathsf{e}^{- \frac{\gamma  \left(\mathbf{g}_{m}^{\top}\mathbf{s} \right)^2}{2} }}
{ \sqrt{ 2 \, \pi}   \, \mathsf{Q}\!\left( -  \sqrt{\gamma}\, b_{m} \, \mathbf{g}_{m}^{\top}\mathbf{s} \right)   }
\label{eq:vector_d_elements},
\end{equation}
$m = 1,2,\ldots, 2M$.  
Note that for any $ \mathbf{s} \in \mathds{R}^{2K}$, 
function $\mathsf{L}(\mathbf{s})$ is an upper bound for the local smoothness parameter
of function $\mathsf{f}(\cdot)$. 

  For the problem in~\eqref{eq:phaseI_proposed_detector},  classic projected 
gradient method  iterates as $\mathbf{s}^{(t+1)} =\mathsf{P}_{\mathcal{B}} \!\left( \mathbf{s}^{(t)} - \eta  \nabla  \mathsf{f}(\mathbf{s}^{(t)})\right)$
 until convergence,  where $\mathsf{P}_{\mathcal{B}}(\cdot)$ is the projector operator onto the set $\mathcal{B}$, given by
 \begin{equation}
  [\mathsf{P}_{\mathcal{B}}(\mathbf{s})]_n = \mathsf{sign}(s_n) \, \min \!\left\{|s_n|, 
  \sqrt{\frac{3( \sqrt{\mathtt{Q}} - 1)^2}{2( {\mathtt{Q}}-1)}} \right\},
  \label{eq:projection_on_cal_B}
 \end{equation}
 $n=1,2,\ldots, 2K$,  and $\eta$ is a suitable constant gradient step  size.
 On the other hand, the proposed accelerated projected gradient method: 
(a) exploits the knowledge of local smoothness upper bound $\mathsf{L}(\cdot)$
in the calculation of the gradient step size
 and (b) employs an extra extrapolation step after projection.
 More specifically, the proposed accelerated projected gradient 
 procedure
is shown in Algorithm~\ref{alg:accel_projected_gradient_adaptive_restart}. 

\begin{algorithm}[!t]
\caption{\small Algorithm to solve problem~\eqref{eq:phaseI_proposed_detector}}
{\small \textbf{Input:} $  \mathbf{G},  \mathbf{b}$, $\gamma$ } 
\begin{algorithmic} [1]
\State{\small Pre-compute $ \| \mathbf{G}\|_2^2$  }
\State{\small $t=0:$ Initialize $\beta^{(0)} = 1$, $\mathbf{u}^{(0)} = \mathbf{s}^{(0)} \in \mathds{R}^{2K}$}
\While { \small Stopping criterion is not reached }
\State{\small $\mathsf{L}( \mathbf{u}^{(t)}) =  \| \mathbf{G}\|_2^2  \, \|\mathbf{d}( \mathbf{u}^{(t)} )\|_{\infty}$}
\State{\small $ \mathbf{s}^{(t+1)} = \mathsf{P}_{\mathcal{B}}\!\left(   \mathbf{u}^{(t )} -
\frac{1}{\mathsf{L}( \mathbf{u}^{(t)})} \nabla  \mathsf{f}(\mathbf{u}^{(t)}) \right)  $ }
\State{\small $\beta^{(t+1)}  = \frac{1 + \sqrt{1 + 4 \left(\beta^{(t)} \right)^2}}{2}$  }
\State{\small$ \mathbf{u}^{(t+1)} = \mathbf{s}^{(t+1)}+ \frac{\beta^{(t)} - 1}{\beta^{(t+1)} }  \left(\mathbf{s}^{(t+1)} -  \mathbf{s}^{(t)}\right)$}
\If {\small $  \nabla \mathsf{f}(\mathbf{u}^{(t)} )^{\top}
\left(\mathbf{s}^{(t+1)} - \mathbf{s}^{(t)}\right)  > 0  $ }
\State{\small $\beta^{(t+1)} = 1$,   $  \mathbf{u}^{(t+1)} = \mathbf{s}^{(t+1)} $}
  \EndIf
\State{\small $t:=t+1$}
  \EndWhile
\end{algorithmic}
{\small \textbf{Output:}  $\breve{\mathbf{s}}^{({\rm I})} = \mathbf{s}^{(t)}  $}
\label{alg:accel_projected_gradient_adaptive_restart}
\end{algorithm}

At line (4), the upper bound of local smoothness parameter of $\mathsf{f}(\cdot)$, $\mathsf{L}(\cdot)$, is calculated,
exploiting the fact that  $\| \mathbf{G}\|_2^2$ can be precomputed.
At line  (5), a projected gradient step is  applied, where the gradient step size harnesses 
the knowledge of  $\mathsf{L}(\cdot)$ at the current point.
Lines (6) and (7) calculate the optimal interpolation parameter 
$\beta^{(t+1)}$  \cite{Bub:15} and apply the interpolation step between the current and the previous points, respectively.
Since function $\mathsf{f}(\cdot)$ is smooth, executing   
lines (4) to (7) iteratively  until convergence,
an $\epsilon$-optimal solution  can be found
(a neighborhood of the optimal solution with diameter $\epsilon \ll 1$) 
using at most $\mathcal{O}(1/\sqrt{\epsilon})$ iterations \cite{Bub:15}.
 An adaptive restart mechanism  (at lines 8--10)
is  also utilized \cite{ODonCan:15} in order to further speed up the convergence rate,
requiring also at most $\mathcal{O}(1/\sqrt{\epsilon})$ iterations to reach an 
$\epsilon$-optimal solution \cite{GisBo:14}.
The algorithm terminates either if  quantity  $\frac{\| \mathbf{s}^{(t+1)} - \mathbf{s}^{(t)}\|_2}{\|\mathbf{s}^{(t)}\|_2}$
is below a prescribed precision  or if a maximum number of iterations is reached.
{In contrast, the projected gradient scheme  with constant step size, 
converges to an optimal solution after $\mathcal{O}(1/\epsilon)$ iterations, which  
is much larger than the proposed  $\mathcal{O}(1/\sqrt{\epsilon})$, especially for small $\epsilon$.}

The calculation of $ \| \mathbf{G}\|_2^2$ requires $\mathcal{O}(K^2 \, M)$ arithmetic operations.
The per iteration complexity of the proposed algorithm is $\mathcal{O}( K \, M)$ due to the evaluation of
$   \nabla  \mathsf{f}(\mathbf{u}^{(t)}) $ and $\mathbf{d}(\mathbf{u}^{(t)})$  
at lines 4 and 5, respectively. In the worst case, the  algorithm iterates
{$I_{\rm max} \approx 1/\sqrt{\epsilon}$  times to find an $\epsilon$-optimal solution, 
requiring  computational cost $\mathcal{O}( \frac{1}{\sqrt{\epsilon}} \,  K \, M )$. 
Thus, the overall  computational cost for  Algorithm~\ref{alg:accel_projected_gradient_adaptive_restart}
is $\mathcal{O}(K \, M  ( 1/\sqrt{\epsilon} + K ) )$.}

As the elements of the $\epsilon$-optimal solution vector $\breve{\mathbf{s}}^{({\rm I})}$ are soft estimates that do not 
necessarily belong to the $\sqrt{\mathtt{Q}}$-PAM constellation set,
after  the execution of 
Algorithm~\ref{alg:accel_projected_gradient_adaptive_restart}   a nearest neighbor rule is employed,
by  projecting  each element of $\breve{\mathbf{s}}^{({\rm I})}$, $\breve{{s}}_n^{({\rm I})}$, to the constellation set $\mathcal{S}$, i.e.,
\begin{equation}
 \widehat{{s}}_n^{({\rm I})} = \arg \min_{s \in \mathcal{S}} | \breve{{s}}_n^{({\rm I})} - s |, ~n=1,2,\ldots, 2K.
 \label{eq:nearest_neighbor_projection}
\end{equation}
Note that $\widehat{\mathbf{s}} ^{({\rm I})} \in \mathcal{S}^{2K}$ and also,
by the properties of ML detector,   $\mathsf{f}(\widehat{\mathbf{s}}^{({\rm I})}) \geq \mathsf{f}(\widehat{\mathbf{s}}^{\rm ML})$ holds.

\subsection{Phase II: Refinement and Multiuser Detection}
\label{subsec:phase_II_detection}

In the second phase (phase II) of the proposed algorithm we apply a refinement step 
to further improve detection performance. First, the vector of absolute residuals  is formed
as
\begin{equation}
z_n \triangleq | \breve{{s}}_n^{({\rm I})} - \widehat{{s}}_n^{({\rm I})}|, ~n=1,2,\ldots, 2K.
\label{eq:vector_residuals}
\end{equation}
The elements of vector $\mathbf{z}$ express the absolute mismatch of the soft-decision estimates and 
the projected estimates of phase I. Intuitively, the smaller the value of a $z_n$ is,
the more  reliable is the estimate for $\breve{{s}}_n^{({\rm I})}$, in the sense that 
$\widehat{{s}}_n^{({\rm I})} = \widehat{{s}}_n^{{\rm ML}}$ with high probability.

After forming vector  $ \mathbf{z}$,  we choose  the $R$ {largest} elements of $ \mathbf{z}$.
Parameter $R$ is a refinement parameter, determining how many  elements 
of estimated vector $ \widehat{\mathbf{s}}^{({\rm I})}$  
and ML vector $ \widehat{\mathbf{s}}^{{\rm ML}}$ may be different.
Refining the decision on  the elements  of decision vector
 $\widehat{\mathbf{s}}^{({\rm I})}$, corresponding to       the indexes of the 
 $R$   {largest}  (less-reliable) elements of vector $ \mathbf{z}$,
 can in principle boost the SER performance of the detector.
Let us denote $\mathcal{J}_{\rm r} \subset \{1,2,\ldots,2K\}$ 
the set of indexes associated with the  $R$ largest elements of 
  residual vector $\mathbf{z}$. 
For each unreliable residual element, i.e., $n \in \mathcal{J}_{\rm r}$, the second closest symbol 
 from the $\sqrt{\mathtt{Q}}$-PAM constellation is obtained through
 \begin{equation}
 \widehat{{s}}_n^{({\rm II})} =  \arg  \min_{s \in \mathcal{S} \backslash \widehat{{s}}_n^{({\rm I})} }  | \breve{{s}}_n^{({\rm I})} - s |, ~n \in \mathcal{J}_{\rm r}.
 \end{equation}
 The  two closest points of $\mathcal{S}$ to the  soft-decision estimate $\breve{{s}}_n^{({\rm I})}$,
 i.e., $ \left \{  \widehat{{s}}_n^{({\rm I})} , \widehat{{s}}_n^{({\rm II})} \right\} $,
 are the refined candidate decision estimates of  symbol $s_n$,  for $n \in \mathcal{J}_{\rm r}$.
On the other hand,  for  $n\in \{1,2,\ldots,2K \} \backslash \mathcal{J}_{\rm r}$,
only the closest point of $\mathcal{S}$ to the soft-decision estimate $\breve{{s}}_n^{({\rm I})}$ (i.e.,
only $   \{  \widehat{{s}}_n^{({\rm I})} \} $, obtained from \eqref{eq:nearest_neighbor_projection}), constitutes the 
single candidate  decision estimate  for symbol  $s_n$.
Combining the above, the refined symbols' codebook can be mathematically expressed as
\begin{equation}
 \mathcal{W}_{\rm r} = \bigotimes_{n \notin \mathcal{J}_{\rm r}} \left \{  \widehat{{s}}_n^{({\rm I})}   \right\} \times
 \bigotimes_{n \in \mathcal{J}_{\rm r}} \left \{  \widehat{{s}}_n^{({\rm I})} , \widehat{{s}}_n^{({\rm II})} \right\},
\end{equation}
forming a set of $2^R$ $2K$-dimensional $\sqrt{\mathtt{Q}}$-PAM symbol sequences.
After forming  the refined  symbols' codebook, the final detector 
is given by
\begin{equation}
\widehat{\mathbf{s}} = \arg \min_{\mathbf{s} \in \mathcal{W}_{\rm r}} \left \{  -\sum_{m=1}^{2M} \mathsf{ln}\,\mathsf{Q}\left( -  \sqrt{\gamma}\,
b_{m} \, \mathbf{g}_{m}^{\top} \,  \mathbf{s} \right) \right \}.
\label{eq:prop_detector}
\end{equation}
The total computational cost to evaluate the  detection  rule  in~\eqref{eq:prop_detector}
is $\mathcal{O} (M\,K \,2^R ) $. {After obtaining  $\widehat{\mathbf{s}}$, BS reconstructs the transmitted complex   $\mathtt{Q}$-QAM symbols  from all
$K$ UEs  as: $\widehat{x}_k = \widehat{s}_k + \mathsf{j}\widehat{s}_{K+k}$, $k=1,2\ldots,K$.}

\subsection{Remarks}
\label{subsec:remarks}

The overall computational cost for the end-to-end multiuser detection procedure,
described in phase I and phase II above, scales with {$\mathcal{O} (M\,K \,(2^R + 1/\sqrt{\epsilon} + K )) $},
which is not exponential on the number of UEs, depending exponentially on the refinement parameter.
The latter controls the accuracy versus complexity trade-off. 
In the studied simulation setups, we found that 
the proposed detector can attain very close SER performance to the ML detector  even for small values of  parameter $R$.

\section{Numerical Results}
\label{sec:num_results}

For the studied simulation setups,  
 the SER performance of the following schemes is studied: (i) proposed two-phase detector, 
 {(ii) the detector in \cite{ChMoHe:16} (two-stage nML), (iii) the detector in \cite{RisPerLar:14} (1-bit ZF), implementable only for 4-QAM, (iv) the Bussgang linear minimum mean-squared error (BLMMSE) detector, that
estimates a soft-decision version of transmitted vector $\mathbf{x}$ using the framework presented in \cite{LiTaSeMeSw:17}
and the elements of the outcome are projected to the $\mathtt{Q}$-QAM constellation set, and  (v) the ML detector. 
   The two-stage  detector in  \cite{ChMoHe:16} requires   $\mathcal{O} (M\,K ( 1/\epsilon) )$  arithmetic operations
 for first stage   plus    $\mathcal{O} (M\,K \, 4^K)$   arithmetic operations for the second stage, using  common  neighborhood parameter $c$ for  all sets in  \cite[Eq.~(38)]{ChMoHe:16}; in contrast to the proposed scheme, the required  complexity
in \cite{ChMoHe:16} is exponential on the number of UEs. 
The computational cost of  BLMMSE detector is $\mathcal{O}(M^2(M+K))$, while
 the simple linear detector in \cite{RisPerLar:14} requires $\mathcal{O} (M\,K^2)$ arithmetic operations.
The proposed detection technique employs   $R = 4$ and $R=6$  for the $4$-QAM and $16$-QAM systems, respectively.}

\begin{figure}[!t]
\centering
\includegraphics[width=0.9\columnwidth]{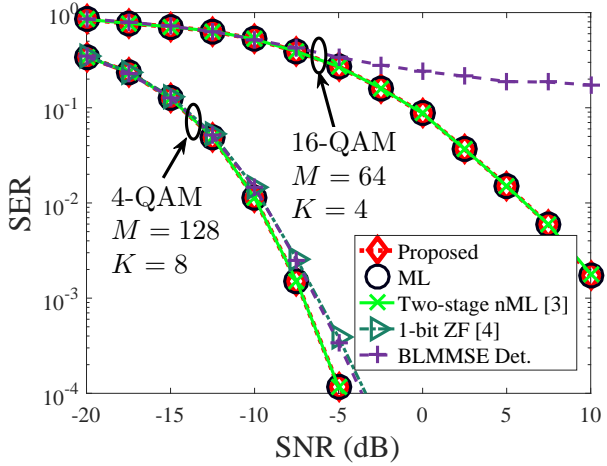}
\caption{SER as a function of SNR.}
\label{fig:SER_vs_SNR}
\end{figure}

In the first simulation study of Fig.~\ref{fig:SER_vs_SNR} 
the SER is plotted as a function of  SNR, using $p_k = 1$ and
$v_k^2 = 1$, for $k=1,2,\ldots, K$, examining also the impact of  
 parameters $K$, $M$, and QAM modulation order, $\mathtt{Q}$.
 For the  $4$-QAM and $16$-QAM systems, the SER of the proposed  and   the two-stage nML detectors 
coincide with the  SER of the ML  detector. The  
ZF 1-bit detector works only for $4$-QAM, while for $16$-QAM the resulting SER is larger than $0.5$;
the algorithm is  computationally cheap but   its SER performance compared to 
the other detectors is worse, especially for high SNR. 
  The BLMMSE detector slightly outperforms ZF 1-bit detector 
and offers slightly worse SER than near ML detectors for the 4-QAM system, while
for 16-QAM system its SER cannot drop below $10\%$.
The proposed detector achieves near optimal performance requiring significantly less  
computational cost compared to the two-stage nML  detector.


\begin{figure}[!t]
\centering
\includegraphics[width=0.9\columnwidth]{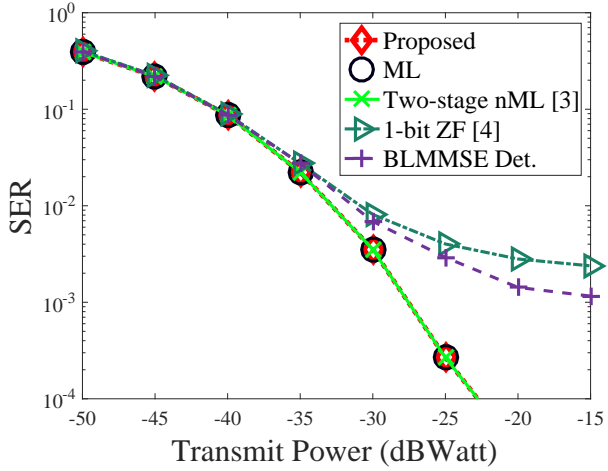}
\caption{SER as a function of transmit power.}
\label{fig:SER_vs_tp}
\end{figure}

In the next simulation setup of Fig.~\ref{fig:SER_vs_tp} we consider  a BS with $M=150$ 
antennas, placed at $[0~0~100]^{\top}$, and $K=8 $ UEs  transmitting 4-QAM symbols,
that are randomly placed  around the BS. 
 The average SER performance of the UEs  
 is examined as a function of  the UE  transmit power,  using $p_k = p$, for $k=1,2,\ldots, K$.
The following path-loss model is considered: 
$v_{k}^2 = (\lambda / 4 \pi )^2 \, (d_{k} / d_0)^{\nu}$, with $d_0 = 100$ meters,
$\nu = 3.2$,  $\lambda = 0.15$, where $d_{k}$ denotes the distance from the $k$th UE
to the BS. The noise power was set $\sigma^2 = -130$ dBWatt.

 In this asymmetric multiuser setting, the SER of 1-bit ZF detector
   saturates after $p = -20$ dBWatt transmit power. BLMMSE detector slightly outperforms 1-bit ZF
and its SER also saturates after $p = -20$ dBWatt. 
The saturation effect stems from the fact that the channel matrix is  ill-conditioned and at the  high-power
regime,    BLMMSE and 1-bit ZF detectors  may offer some erroneous detection decisions due to the 
required channel inversion.
Both BLMMSE and 1-bit ZF detectors
offer similar SER with nML detectors, but beyond $p = -30$ dBWatt their performance becomes worse.
 On the other hand, the proposed 
and the two-stage nML detectors achieve similar SER with  the optimal ML detector.

%

%
{
\section{Conclusion}
\label{sec:conclusion}

In this work a two-phase 
detection algorithm is proposed for uplink multiuser  massive MIMO
systems employing single-bit ADCs. The algorithm achieves 
the SER performance of the ML detector and manages to significantly reduce  the computational cost of the
nML detectors in prior art.

}



\end{document}